# Anomalous phase shift of Shubnikov - de Haas oscillations in HgTe quantum well with inverted energy spectrum

Neverov V.N.[1*], Klepikova A.S.[1], Bogolubskii A.S.[1], Gudina S.V.[1], Turutkin K.V.[1], Shelushinina N.G.[1], Yakunin M.V.[1,2], N.N. Mikhailov[3] and S.A. Dvoretsky[3]

[1]*M.N. Miheev Institute of Metal Physics of Ural Branch of Russian Academy of Sciences, 18 S. Kovalevskaya Str., Ekaterinburg 620990, Russia*

[2]*Ural Federal University, 19 Mira Str., Ekaterinburg 620002, Russia*

[3]*A.V. Rzhanov Institute of Semiconductor Physics of Siberian Branch of Russian Academy of Sciences, 13 Lavrentyev Ave., Novosibirsk 630090, Russia*

*neverov@imp.uran.ru

The results of the longitudinal and Hall magnetoresistivity measurements in the Shubnikov - de Haas oscillation regime for the HgCdTe/HgTe/HgCdTe heterostructures with a wide (20.3 nm) HgTe quantum well are presented. An anomalous phase shift of magneto-oscillations is detected: in the region of spin-unsplit peaks the longitudinal resistivity maxima are located at even filling factor numbers in contradiction with a conventional situation in 2D systems.

It is shown that the observed features are associated with the inverted nature of the spectrum in the investigated quantum well with the electron-type conduction along the size-quantized subband $H1$ of HgTe band $\Gamma_8$, for which the spin splitting is comparable to (and even greater than) the orbital one.

The results obtained are compared with the phase shift effects of both magneto-oscillations and the plateau of the quantum Hall effect in monolayer graphene.

## Introduction

It is well-known [1] that both HgTe and CdTe bulk materials have the zinc-blende lattice structure where the actual extrema in the bands are close to the $\Gamma$-point of the Brillouin zone, and they are the $s$-type band ($\Gamma_6$) and the $p$-type band, which is split to a $J = 3/2$ -band ($\Gamma_8$) and a $J = 1/2$ -band ($\Gamma_7$) by spin-orbit coupling.

CdTe has a "normal" band order with $\Gamma_6$ conduction band, $\Gamma_8$ and $\Gamma_7$ valence bands. The highest valence band $\Gamma_8$ is separated from the conduction band by a large energy gap $\varepsilon_g = 1.6$ eV ($\varepsilon_g \equiv E(\Gamma_6) - E(\Gamma_8)$).

In a bulk HgTe, the $\Gamma_8$ band, which "normally" forms the valence band, is above the $\Gamma_6$ band due to relativistic effects [1], that is indicated by a negative sign of the energy gap: $\varepsilon_g = -300$ meV. The light-hole subband of the bulk $\Gamma_8$ band becomes the conduction band and the heavy-hole subband becomes the first valence band. Based on this unusual sequence of the $\Gamma_6$ and $\Gamma_8$ states, such a band structure is called "inverted".

When Cd(Hg)Te/HgTe/Cd(Hg)Te heterostructures with HgTe quantum well (QW) are grown [2], the quantum confinement gives rise to the "normal" sequence of subbands for a thin enough QW layer, similar to CdTe, i.e., the bands with primarily $\Gamma_6$ symmetry are the conduction subbands and the $\Gamma_8$ bands contribute to the valence subbands.

As the QW thickness is increased, the material looks more and more like HgTe and for wide enough QW layers the band structure becomes "inverted". This happens when the QW width, $d_{QW}$, exceeds a critical value $d_c \approx 6.3$ nm. For the inverted energy spectrum of the HgTe QW, the first size-quantized heavy-hole subband $H1$ becomes the lowest conduction subband [2]. Formally, the $H1$ sublevel belongs to the heavy hole branch of the $\Gamma_8$ band with the $z$-component of the total momentum $J_z = \pm 3/2$, however, the theory [3,4] predicts an electron-like effective mass for it.

We present a study of the Shubnikov - de Haas oscillations (SHO) in a 20.3-nm-wide HgTe QW with conduction carried out by the electrons of the $H1$ size-quantized subband with an extremely small effective mass $m_c/m_0$ and a large value of the $g$-factor.

**Theoretical background**

Let us compare the orbital (cyclotron), $\Delta_c = eB/m_c$, and spin (Zeeman), $\Delta_s = g\mu_B B$, splittings of the $H1$ Landau levels (LLs). For the effective mass at the bottom of $H1$ subband (at $k_\parallel = 0$), $m_c$, the theory in spherical approximation gives [5-8]:

$$m_0/m_c = \gamma_1 + \gamma, \text{ i.e.}$$

$$\Delta_c = (\gamma_1 + \gamma)\hbar\omega_0. \qquad (1)$$

The energetic distance between two sublevels with $J_z = \pm 3/2$ can be written in the following form [5-8]:

$$\Delta_s = \varepsilon_{3/2} - \varepsilon_{-3/2} = 3\kappa\hbar\omega_0, \qquad (2)$$

i.e. the theoretical value of the effective $g$-factor, $g = 6\kappa$, and

$$\Delta_s - \Delta_c = [3\kappa - (\gamma_1 + \gamma)]\hbar\omega_0. \tag{3}$$

Here, $\gamma_1$, $\gamma$ and $\kappa$ are the Luttinger parameters, $\omega_0 = eB/m_0$ being the free electron cyclotron frequency and $\mu_B = e\hbar/2m_0$ – the Bohr magneton. For a set of HgTe parameters ($\gamma_1 = 12.8$, $\gamma = 8.4$ and $\kappa = 10.5$) [9], we have $3\kappa > (\gamma_1 + \gamma)$, and, at least for small $k_\parallel$, the spin splitting is *greater* than the orbital one, $\Delta_s > \Delta_c$.

Thus, the theoretical values of parameters at the heavy-hole 2D-subband bottom $H1$ for HgTe QW are $g \cong 60$ and $m_c/m_0 = 0.047$. But the experimental estimations of the cyclotron and Zeeman energy from the activation analysis in high magnetic fields as well as the estimations of the cyclotron energy from the weak-field SHO may contradict these values of $g$ and $m_c/m_0$ due to a mixed nature of heavy-hole subbands at finite $k_\parallel$ that leads to a pronounced nonlinearity of the $B$ dependence of the LLs in the inverted-band regime (see, for example, [10]).

Thus, in Ref. [11], the activation analysis of experimental magnetoresistivity traces was used as a quantitative tool to probe inter-LL distances for HgCdTe/HgTe/HgCdTe system with HgTe quantum well width of 20.3 nm. The activation energies, $\Delta_\nu$, were determined from the temperature dependence of the longitudinal resistivity in the regions of quantized Hall plateaus (for the filling factors $\nu$ of 2 and 3).

Using the experimental values of $\Delta_2$ and $\Delta_3$ we get the following estimates for the parameters for the $H1$ subband: $m_c/m_0 = 0.037 \pm 0.005$ and $g = 75 \pm 5$. Note that such a large $g$-factor value for a HgTe QW with inverted band structure is due to the *p*-type nature of states in the $\Gamma_8$ size-quantized subband $H1$ with the total angular momentum $J = 3/2$ and *z*-projections of the "quasi-spin" $J_z = \pm 3/2$, in contrast to the standard situation with $S_z = \pm 1/2$. The renormalization of the $g$-factor with respect to its bottom value $g = 6\kappa$ is associated with magnetically induced mixing of the heavy-hole states with the light-hole ones [8, 12].

**Experimental results and discussion**

We investigate the SHO-regime in the longitudinal, $\rho_{xx}$, and Hall, $\rho_{xy}$, resistivities in magnetic fields $B$ up to 2.5 T at temperatures $T = (2$–$10)$ K for the HgCdTe/HgTe/HgCdTe heterostructure with the HgTe quantum well width of 20.3 nm, grown on the (013) GaAs substrate and symmetrically modulation doped with In. In the studied QW with an inverted band structure, the conduction is carried out by electrons in the size-quantized subband $H1$ with an extremely small effective mass $m_c/m_0$ and a large value of $g$-factor [13-15]. Electron concentration $n_s = 1.5 \times 10^{15}$ m$^{-2}$ and mobility $\mu = 22$ m$^2$/Vc.

The results for measured $\rho_{xx}(B)$ and $\rho_{xy}(B)$ in the investigated sample are presented in Fig. 1 for magnetic fields $B < 2$ T at $T = 2$ K with the features of quantum Hall effect (QHE) near 2 T. SHO regime corresponds to the filling factors 4 and above with the spin-unsplit peaks for $\nu > 7$, $B < B^* \cong 1$T (also see the inset).

*The anomalous phase shift of the SdH oscillations*

We note an important characteristic property in the region of doubly degenerate peaks of $\rho_{xx}$ ($B < B^*$): the peaks of $\rho_{xx}(B)$ are observed for *even* filling factors and minima correspond to *odd* $\nu = (7 - 21)$ (see Fig. 1). This is in contradistinction to the conventional situation for spin-degenerate case in 2D systems.

If we assume that the spin splitting is comparable with the cyclotron splitting, $g\mu_B B \gtrsim \hbar\omega_c$ (and there are reasons for this because of the ratio of $\gamma$-parameters of the $\Gamma_8$ band in HgTe; see details above), then we have the scheme of LLs and the corresponding density of states in a quantizing magnetic field presented in Fig. 2.

The region of spin-unsplit peaks at $B < B^*$ then corresponds to the condition $|g\mu_B B - \hbar\omega_c| < \Gamma$ with $\Gamma$ being the Landau level broadening. In this case, the two-fold degeneracy of the $\rho_{xx}$ peaks are due to the proximity of the energies for the adjacent Landau levels with oppositely directed (*antitropic*) spins, $N \downarrow$ and $(N-1) \uparrow$ (see Fig. 2b). Thus, we have a system of doubly degenerate Landau levels with an "extra" non-degenerate level for $N = 0\downarrow$ (see Fig. 2a).

The situation is similar to a monolayer graphene [16-18] with the difference being that peaks in graphene are fourfold degenerate. In graphene, the lowest LL with $N = 0$ appears at $E = 0$ and each level, including zero LL, has 4-fold degeneracy (2 spins× 2 pseudospins). The origin of the latter is associated with the presence of two carbon sublattices. Zero LL in the Dirac-like spectrum of graphene is half-populated by electrons (2-fold degeneracy) and half-populated by holes (2-fold degeneracy). See, e.g., Fig.2c in Ref. [17] or Fig 1(d) in Ref. [18]: LL ladder in monolayer graphene for explanation of the anomalous phase of the magneto-oscillations.
If you focus on only one (e.g., electronic) branch of the spectrum, then this property implies that the degeneracy for $N = 0\downarrow$ is half of that for any other $N$. This explains the "half-integer" QHE plateau [16]:

$$\rho_{xy}^{-1} = g_s(i + 1/2)\,e^2/h \qquad (4)$$

at filling factors $\nu = g_s(i + 1/2)$ (with $g_s = 4$ and $i = 0, 1, 2 \ldots$ for graphene) and, at the same time an anomalous phase of SHO (minima in $\rho_{xx}$ correspond to plateau in $\rho_{xy}$ and maxima in $\rho_{xx}$ correspond to transitions between the plateau).

In our system, these considerations are applicable for $g_s = 2$ (in the region where there is no spin splitting, $\nu > 7$) and thus minima in $\rho_{xx}$ should be at the *odd* $\nu = 2(i + 1/2)$ (with $i \geq 3$) and maxima in $\rho_{xx}$ ought to be at the *even* $\nu = 2i$ (with $i \geq 4$) as it is actually observed in the experiment.

We emphasize, however, that while the behavior of $\rho_{xx}(B)$ and $\rho_{xy}(B)$ both in the field of oscillations and in the QHE regime for graphene is due to fundamental physical reasons, the features observed by us have a "technical" origin: it is due to a specific ratio of parameters ($\gamma$ and $\kappa$ parameters) in a specific band ($\Gamma_8$ band) of specific material (HgTe).

In our system, the double degeneracy of levels is removed at fields $B > B$ * that corresponds to the condition $|g\mu_B B - \hbar\omega_c| > \Gamma$, and the LL spectrum becomes "normal": all Landau levels are equally non-degenerate. Therefore, at this range of fields there is no anomalous phase shift, and the conventional QHE with the integer plateaus of $i = 1, 2, 3$ and the peaks of $\rho_{xx}(B)$ at $\nu = 1.5; 2.5;$ and $3.5$ is observed (see [11]).

On the other hand, in graphene, due to a small value of the $g$ -factor ($g \cong 2$ [19]), spin degeneracy is not removed even in the field of 14T where the QHE data are presented in Ref.[16] (see Fig. 4 in [16] with QHE for massless Dirac fermions). Thus, the difference in degree of degeneracy for LL with $N = 0$ (2-fold) and the remaining LLs (4-fold) is also preserved in the QHE regime, which leads to an abnormal phase shift and as a result to the half-integer QHE plateaus:

$$\left(\frac{h}{4e^2}\right)\sigma_{xy} = \pm\frac{1}{2}; \pm\frac{3}{2}; \pm\frac{5}{2}, \text{etc.}$$

*Fan diagram*

Note, that from another viewpoint the phase shift of SdH oscillations and, at the same time, the half-integer QHE in 'ideal' graphene is considered as the manifestation of Berry's phase (a geometric quantum phase of the electron wavefunction) acquired by Dirac fermions moving in magnetic field [20, 21].

For the phase analysis of quantum oscillations, a plot of the location, $1/B$, for *n*-th minimum (maximum) in dependence on its number $n$ ($n + 1/2$) should be constructed (Berry diagram). Then the straight lines, corresponding to the linear fit, intersect the $n$ axis at some point $n = n^*$:

with an integer value of n* for topologically trivial systems and with a half-integer n* for topologically nontrivial one.

It is the kind of graph that is widely used for a direct probe of Berry phase in the magneto-oscillation of systems with a topologically nontrivial energy spectrum, first of all, in graphene, as well as in 2D - or 3D - topological insulators (see [18] and references therein).

Such a graph for a single-layer graphene, adopted from Ref. [18], is presented on the inset of Fig. 3: diagram $1/B$ versus both a filling factor, $\nu$, at top axis and a number of minimum, $n$, at bottom axis (see discussion in [18]) It is seen that the dependence of $1/B$ on the filling factor $\nu$ (defined as $h/(e^2 \rho_{xy})$) extrapolates to zero with $1/B \to 0$, which naturally follows from the definition of $\nu$ as the number of occupied Landau levels:

$$\nu = n_s/n_B \sim 1/B \tag{5}$$

with $n_s$ and $n_B = eB/h$ being the electron concentration and the number of states at a Landau level per unit area, respectively.

To identify a nontrivial phase shift (Berry phase), it is necessary to use the dependence of $1/B_{min}$ on the number $n$ of the $\rho_{xx}$ minimum, regardless of the degree of the peak degeneracy [16-18]. Extrapolation of the straight line $1/B_{min}(n)$ to half integer (!) value $n^* = 1/2$ with $1/B \to 0$ for graphene (see inset on Fig.3) just considered as a manifestation of the non-trivial Berry phase due to the peculiar topology of the graphene band structure with a linear dispersion relation and vanishing mass near the Dirac point.

Fig. 3 shows the dependences of the inverse magnetic fields, $1/B_{min}$, corresponding to the minima of the magnetoresistivity for investigated sample in two versions: in dependence on the filling factor, $\nu$, or in dependence on the number, $n$, of the observed minimum. For our sample, we naturally see on the $1/B_{min}(\nu)$ dependence that a cutoff on the $\nu$ axis is also equal to zero. However, the situation is quite different for the dependence of $1/B_{min}$ on $n$: extrapolating it to $1/B \to 0$ from the region of unsplit peaks ($\nu \geq 7$) yields the limit $n^* = 3.5$, i.e. a half integer (!) cutoff value.

However, the found half-integer value of $n^*$ has nothing to do with the genuine Berry phase for topologically nontrivial systems. Here, it is a simple consequence of the fact that for $B < B^*$ in the system of doubly degenerate Landau levels there is an "extra" non-degenerate level for $N = 0$ due to the special relation $g\mu_B B \cong \hbar\omega_c$ for HgTe QW with an inverted energy spectrum. Thus, though it looks like in graphene, but it is still not the same as that in graphene.

Let us draw attention that the specific value of $n^* = 3.5$ is due to the fact that the spin degeneracy is retained only until $\nu > 7$ ($B < B^*$). For $\nu < 7$ ($B > B^*$), all the LLs are non-degenerate, the concepts of $\nu$ and $n$ coincide, and we have a single straight line that goes to zero as $1/B \rightarrow 0$ on the graph of Fig. 3.

**Conclusions**

Reach information concerning the electron spectrum in 2D system may be obtained from a study of Shubnikov - de Haas quantum oscillation effect for longitudinal magnetoresistance. We have investigated the regime of SHO in magnetic fields $B$ up to 2.5T at temperatures $T$ = (2–10) K for HgCdTe/HgTe/HgCdTe heterostructure with HgTe quantum well width of 20.3 nm. In studied QW with an inverted band structure the conduction is carried out by electrons of the size-quantized subband $H$1.

The ratio of the Luttinger parameters ($\gamma, \gamma_1$ and $\kappa$) of the $\Gamma_8$ band in HgTe is such that the spin splitting in the $H$1 subband is comparable to the orbital one, at least for small $k_{//}$. Previously, in Ref. [11], we presented the related features of the LLs spectrum in the regime of the quantum Hall effect ($\omega_c \tau \gg 1$). Studying the magnetoresistance of this system with a relatively large filling factor in intermediate magnetic fields, for $\omega_c \tau > 1$, we have found yet another interesting manifestation of its specificity, namely, the anomalous phase shift of SHO in the region of doubly degenerate peaks.

A large value of the spin splitting, comparable to the cyclotron energy, leads to a system of doubly degenerate LLs with an "extra" non-degenerate level for $N$ = 0↓ in not too strong magnetic fields in our system. This kind of the imbalance in the degree of degeneracy for LL with N = 0 (2-fold) and the remaining LLs (4-fold) for electrons (or holes) in monolayer graphene, as a consequence of the exceptional topology of the graphene band structure, leads to the existence of a non-zero Berry phase in SHO and to a distinctive half-integer quantum Hall effect [16-18].

We emphasize, that in the system studied, the explanation for the LL imbalance and for the corresponding phase shift of the magneto-oscillation has nothing to do with the features of the band structure topology, but is more trivial: a specific ratio of parameters ($\gamma$ and $\kappa$ parameters) in $\Gamma_8$ band of HgTe.

**Acknowledgements**

The research was carried out within the state assignment of Russian Ministry of Science and High education, theme "Electron" and "Function", supported in part by RFBR, projects Nos. 18-02-00172 (experiment), 20-42-660004 (theoretical support).

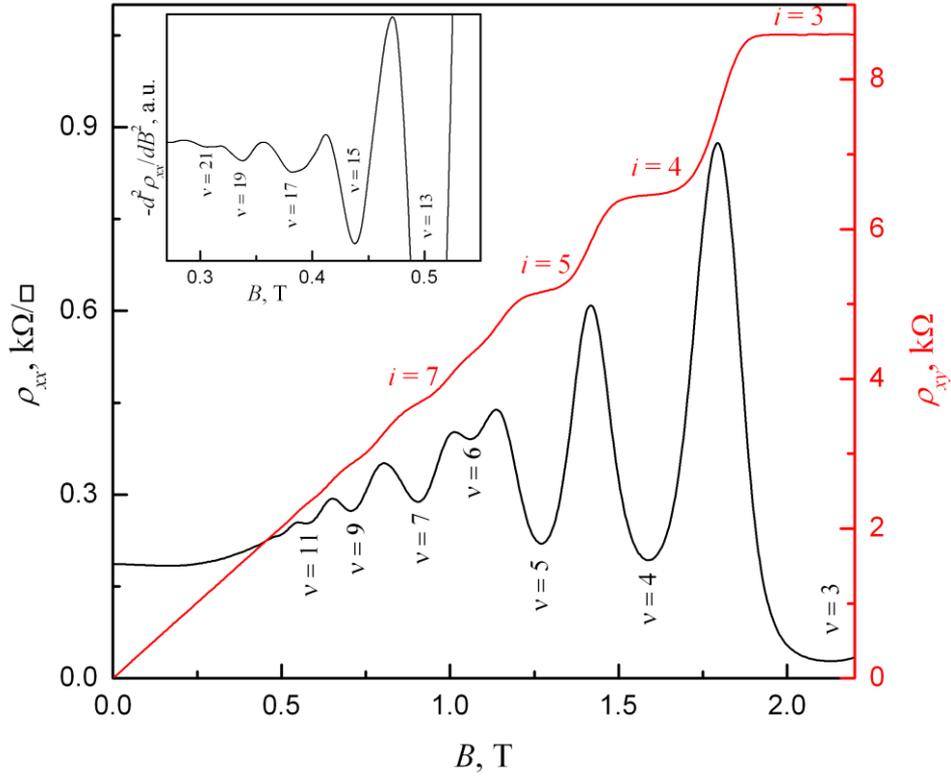

Fig.1 The dependences of longitudinal, $\rho_{xx}$, and Hall, $\rho_{xy}$, magnetoresistivities on the magnetic field at $T$ = 2 K for investigated sample. QHE plateau integer numbers $i = \frac{e^2}{h}\rho_{xy}$ are marked on the curve for $\rho_{xy}(B)$. The indices ν = (3-11) near minima of $\rho_{xx}(B)$ indicate corresponding filling factors. Inset: the second derivative of $\rho_{xx}$ (with a minus sign) for large filling factors ν = (13-21) with corresponding index numbers shown near the minima.

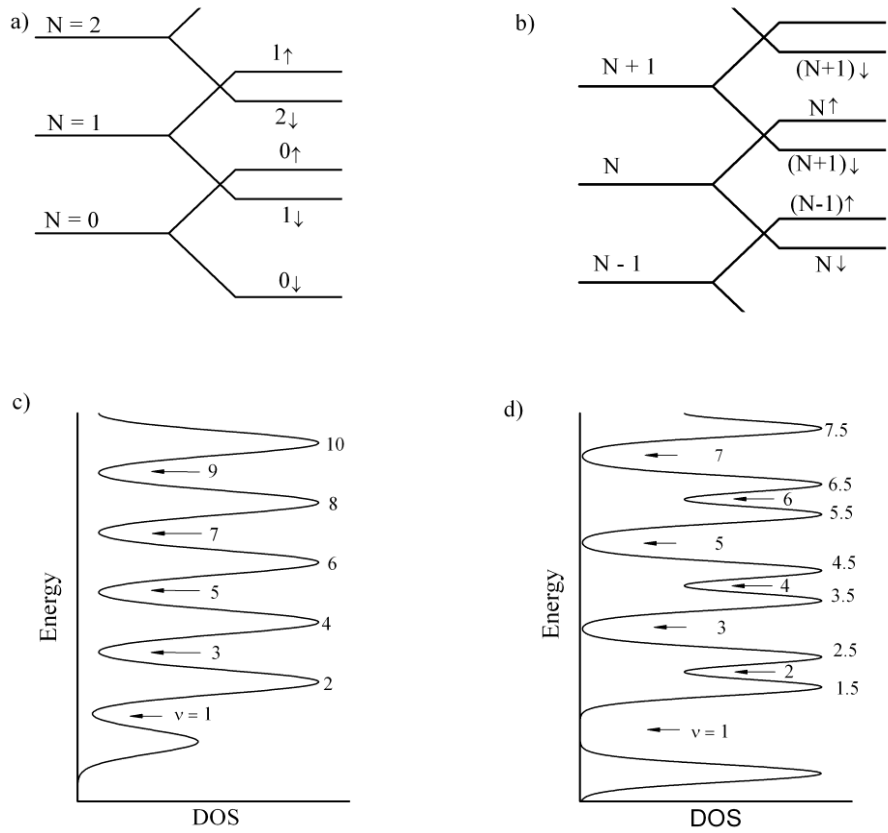

Fig.2. A schematic diagram of the Landau level spectrum for lower LLs (a) and for the general case with the LL index equals $(N-1)$, $N$ and $(N+1)$ (b) in a given magnetic field. The LL indices $N\uparrow$ and $N\downarrow$ are shown next to the LLs. Density of states (DOS) corresponding to above LL scheme without spin splitting, $B < B^*$ (c) and with it, $B > B^*$ (d). The filling factors ν are shown next to the DOS peaks and dips.

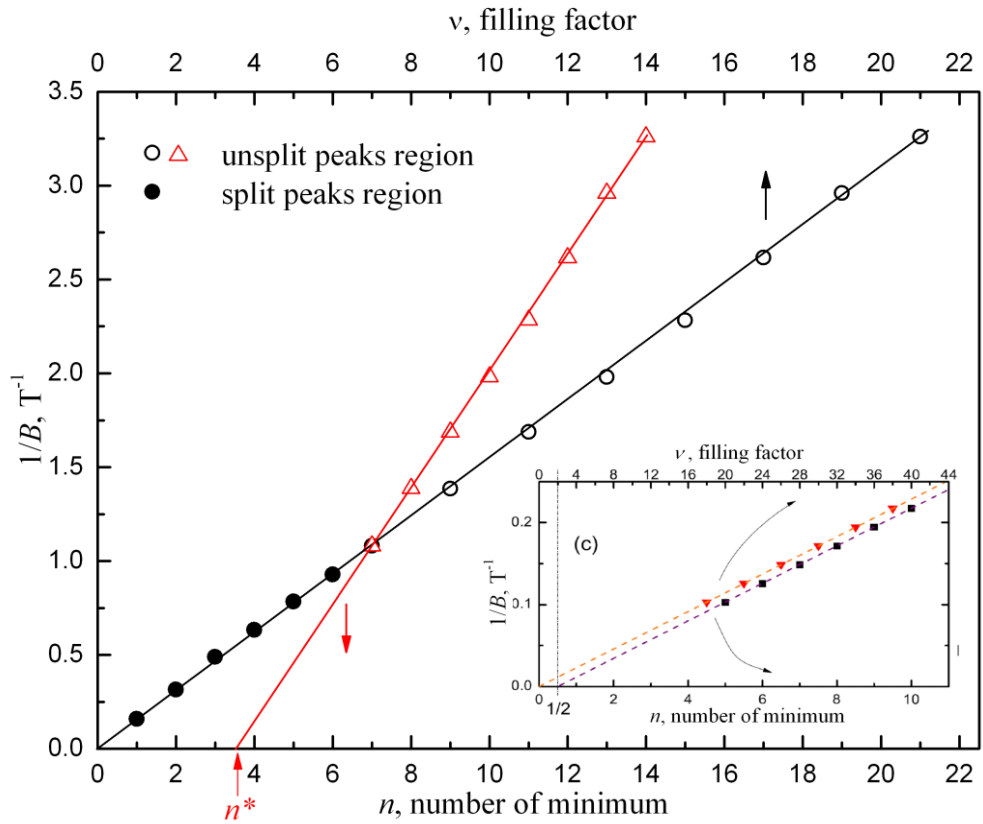

Fig. 3. The inverse magnetic fields $1/B_{min}$ in dependence on filling factor $\nu$ or on number $n$ of the observed magnetoresistivity minimum for investigated sample. The extrapolation of $1/B_{min}(n)$ to $1/B \rightarrow 0$ from the region of unsplit peaks ($\nu \geq 7$) tends to $n^* = (3.55 \pm 0.05)$.

Inset: an example of Berry diagram for monolayer graphene from Ref. [18] (see text).